\documentclass[conference,10pt,a4paper]{IEEEtran}
\usepackage{amsmath}
\usepackage{times}
\usepackage{graphicx,import}
\usepackage{multirow}
\usepackage{tabularx}
\usepackage{makecell}
\usepackage[none]{hyphenat}
\usepackage{float}
\usepackage{subfig}
\usepackage{siunitx}
\usepackage[c]{esvect}
\usepackage{enumitem}
\usepackage{hyperref}

\makeatletter

\def\@maketitle{\newpage
\bgroup\par\addvspace{0.5\baselineskip}\centering%
\ifCLASSOPTIONtechnote
   {\bfseries\large\@IEEEcompsoconly{\sffamily}\@title\par}\vskip 1.3em{\lineskip .5em\@IEEEcompsoconly{\sffamily}\@author
   \@IEEEspecialpapernotice\par{\@IEEEcompsoconly{\vskip 1.5em\relax
   \@IEEEtitleabstractindextextbox{\@IEEEtitleabstractindextext}\par
   \hfill\@IEEEcompsocdiamondline\hfill\hbox{}\par}}}\relax
\else
   \vskip0.2em{\EuMWtitlesize\ifCLASSOPTIONtransmag\bfseries\LARGE\fi\@IEEEcompsoconly{\sffamily}\@IEEEcompsocconfonly{\normalfont\normalsize\vskip 2\@IEEEnormalsizeunitybaselineskip
   \bfseries\Large}\@title\par}\vskip1.0em\par
   \ifCLASSOPTIONconference%
      {\@IEEEspecialpapernotice\mbox{}\vskip\@IEEEauthorblockconfadjspace%
       \mbox{}\hfill\begin{@IEEEauthorhalign}\@author\end{@IEEEauthorhalign}\hfill\mbox{}\par}\relax
   \else
      \ifCLASSOPTIONpeerreviewca
         {\@IEEEcompsoconly{\sffamily}\@IEEEspecialpapernotice\mbox{}\vskip\@IEEEauthorblockconfadjspace%
          \mbox{}\hfill\begin{@IEEEauthorhalign}\@author\end{@IEEEauthorhalign}\hfill\mbox{}\par
          {\@IEEEcompsoconly{\vskip 1.5em\relax
           \@IEEEtitleabstractindextextbox{\@IEEEtitleabstractindextext}\par\hfill
           \@IEEEcompsocdiamondline\hfill\hbox{}\par}}}\relax
      \else
         \ifCLASSOPTIONtransmag
           {\@IEEEspecialpapernotice\mbox{}\vskip\@IEEEauthorblockconfadjspace%
            \mbox{}\hfill\begin{@IEEEauthorhalign}\@author\end{@IEEEauthorhalign}\hfill\mbox{}\par
           {\vspace{0.5\baselineskip}\relax\@IEEEtitleabstractindextextbox{\@IEEEtitleabstractindextext}\vspace{-1\baselineskip}\par}}\relax
         \else
           {\lineskip.5em\@IEEEcompsoconly{\sffamily}\sublargesize\@author\@IEEEspecialpapernotice\par
           {\@IEEEcompsoconly{\vskip 1.5em\relax
            \@IEEEtitleabstractindextextbox{\@IEEEtitleabstractindextext}\par\hfill
            \@IEEEcompsocdiamondline\hfill\hbox{}\par}}}\relax
         \fi
      \fi
   \fi
\fi\par\addvspace{0.0\baselineskip}\egroup}

\def\EuMWtitlesize{\@setfontsize{\EuMWtitlesize}{24}{24pt}}
\def\EuMWauthorsize{\@setfontsize{\EuMWauthorsize}{11}{11pt}}
\def\EuMWaffilsize{\@setfontsize{\EuMWaffilsize}{10}{10pt}}
\def\EuMWcaptionsize{\@setfontsize{\EuMWcaptionsize}{9}{10pt}}
\def\EuMWbibsize{\@setfontsize{\EuMWbibsize}{8}{10pt}}

\def\@IEEEauthorblockNstyle{\EuMWauthorsize\@IEEEcompsocnotconfonly{\sffamily}\@IEEEcompsocconfonly{\large}}
\def\@IEEEauthorblockAstyle{\EuMWaffilsize\@IEEEcompsocnotconfonly{\sffamily}\@IEEEcompsocconfonly{\itshape}\@IEEEcompsocconfonly{\large}}
\def\@IEEEauthordefaulttextstyle{\EuMWauthorsize\@IEEEcompsocnotconfonly{\sffamily}\sublargesize}

\def\thebibliography#1{\section*{\refname}%
    \addcontentsline{toc}{section}{\refname}%
    \EuMWbibsize\@IEEEcompsocconfonly{\small}\vskip 0.3\baselineskip plus 0.1\baselineskip minus 0.1\baselineskip
    \list{\@biblabel{\@arabic\c@enumiv}}%
    {\settowidth\labelwidth{\@biblabel{#1}}%
    \leftmargin\labelwidth
    \advance\leftmargin\labelsep\relax
    \itemsep \IEEEbibitemsep\relax
    \usecounter{enumiv}%
    \let\p@enumiv\@empty
    \renewcommand\theenumiv{\@arabic\c@enumiv}}%
    \let\@IEEElatexbibitem\bibitem%
    \def\bibitem{\@IEEEbibitemprefix\@IEEElatexbibitem}%
\def\newblock{\hskip .11em plus .33em minus .07em}%
\ifCLASSOPTIONtechnote\sloppy\clubpenalty4000\widowpenalty4000\interlinepenalty100%
\else\sloppy\clubpenalty4000\widowpenalty4000\interlinepenalty500\fi%
    \sfcode`\.=1000\relax}

\long\def\@makecaption#1#2{%
\ifx\@captype\@IEEEtablestring%
\par\@IEEEtabletopskipstrut
\else
\@IEEEfigurecaptionsepspace
\fi
\setbox\@tempboxa\hbox{\normalfont\footnotesize {#1.}\nobreakspace\nobreakspace #2}%
\ifdim \wd\@tempboxa >\hsize%
\setbox\@tempboxa\hbox{\normalfont\footnotesize {#1.}\nobreakspace\nobreakspace}%
\parbox[t]{\hsize}{\normalfont\footnotesize\noindent\unhbox\@tempboxa#2}%
\else
\ifCLASSOPTIONconference \hbox to\hsize{\normalfont\footnotesize\hfil\box\@tempboxa\hfil}%
\else \hbox to\hsize{\normalfont\footnotesize\box\@tempboxa\hfil}%
\fi\fi
\ifx\@captype\@IEEEtablestring%
\@IEEEtablecaptionsepspace
\else
\fi}

\newlength\tablecaptiontotableskip
\newlength\figuretocaptionskip
\def\@IEEEfigurecaptionsepspace{\vskip\figuretocaptionskip\relax}%
\def\@IEEEtablecaptionsepspace{\vskip\tablecaptiontotableskip\relax}%

\def\abstract{\normalfont%
\@IEEEabskeysecsize\bfseries\textit{\abstractname}\,\bfseries\textit{---}\,%
\@IEEEgobbleleadPARNLSP}%

\def\IEEEkeywords{\normalfont%
\@IEEEabskeysecsize\bfseries\textit{\IEEEkeywordsname}\,\bfseries\textit{---}\,%
\@IEEEgobbleleadPARNLSP}%
\def\endIEEEkeywords{\relax\vspace{0.67ex}%
\par\if@twocolumn\else\endquotation\fi%
\normalsize\normalfont}%

\DeclareRobustCommand*{\EuMWauthorrefmark}[1]{\raisebox{0pt}[0pt][0pt]{\textsuperscript{\footnotesize{#1}}}}%
\def\@IEEEauthorblockNtopspace{0ex}
\def\@IEEEauthorblockAtopspace{1mm}
\def\IEEEkeywordsname{Keywords}
\def\subsubsection{\@startsection{subsubsection}{3}{\z@}{1.5ex plus 1.5ex minus 0.5ex}%
{0.7ex plus .5ex minus 0ex}{\normalfont\normalsize\itshape}}%
\newlength{\CPheadmatchindent}%
\def\@seccntformat#1{\hbox to\CPheadmatchindent{\csname the#1dis\endcsname}\hskip 0.1em \relax}
\IEEEilabelindentA \parindent
\IEEEilabelindent \IEEEilabelindentA
\IEEEelabelindent \parindent
\IEEEdlabelindent \parindent
\IEEElabelindent \parindent
\makeatother

\begin{document}
\raggedbottom
%
%
%
\title{An Efficient yet High-Performance Method for Precise Radar-Based Imaging of Human Hand Poses}
%
%
\author{%
\IEEEauthorblockN{%
Johanna Br{\"a}unig\EuMWauthorrefmark{\#1},
Vanessa Wirth\EuMWauthorrefmark{*}, 
Marc Stamminger\EuMWauthorrefmark{*},
Ingrid Ullmann\EuMWauthorrefmark{\#}, 
Martin Vossiek\EuMWauthorrefmark{\#}
}
\IEEEauthorblockA{%
\EuMWauthorrefmark{\#}Institute of Microwaves and Photonics (LHFT), Friedrich-Alexander-Universit{\"a}t Erlangen-N{\"u}rnberg, Germany\\
\EuMWauthorrefmark{*}Chair of Visual Computing (LGDV), Friedrich-Alexander-Universit{\"a}t Erlangen-N{\"u}rnberg, Germany \\
\EuMWauthorrefmark{1}johanna.braeunig@fau.de\\
}
}
%
\maketitle
%
%
\begin{abstract}
Contactless hand pose estimation requires sensors that provide precise spatial information and low computational complexity for real-time processing. Unlike vision-based systems, radar offers lighting independence and direct motion assessments. Yet, there is limited research balancing real-time constraints, suitable frame rates for motion evaluations, and the need for precise 3D data. To address this, we extend the ultra-efficient two-tone hand imaging method from our prior work to a three-tone approach. Maintaining high frame rates and real-time constraints, this approach significantly enhances reconstruction accuracy and precision. We assess these measures by evaluating reconstruction results for different hand poses obtained by an imaging radar. Accuracy is assessed against ground truth from a spatially calibrated photogrammetry setup, while precision is measured using 3D-printed hand poses. The results emphasize the method’s great potential for future radar-based hand sensing.
\end{abstract}
%
%
\section{Introduction}
Contactless hand shape and pose reconstruction, crucial for applications like human-computer interaction (HCI), augmented/virtual reality, robotics, and medicine, mostly relies on vision-based approaches utilizing RGB, depth, or RGB-D cameras \cite{Huang.2021}, which have the drawback of lighting dependency. In this context, radar sensors are a robust alternative, providing accurate spatial information even under difficult lighting conditions. Besides that, radar systems can directly measure the radial speed of movements, which is beneficial in hand tracking applications. Due to commonly employed radar hardware incorporating low lateral resolution, radar-based hand sensing applications are often limited to the recognition of characteristic motion patterns \cite{Ahmed.2021}. For tasks requiring the resolution of different fingers, such as static hand gesture recognition \cite{Schuessler.2024} or precise shape and pose estimation, the lateral resolution needs to be increased. Traditional radar near-field imaging techniques, commonly utilizing multiple input multiple output (MIMO) radars, rely on high signal bandwidths and a high number of frequency steps or samples, leading to low frame rates, unsuitable for simultaneous Doppler measurements~\cite{Ahmed.2021b}. Furthermore, state-of-the-art (SOTA) approaches often involve computationally expensive reconstruction algorithms, hindering real-time evaluation crucial for HCI applications. For this reason, in \cite{Braunig.2023} we proposed an ultra-efficient approach to measure and reconstruct the 3D surface of the human hand using microwave near-field imaging. The proposed approach reduces measurement acquistion times as only two continuous wave (CW) frequencies need to be transmitted, opening the way towards simultaneous Doppler and imaging measurements. Furthermore, by implementing a novel signal processing concept that combines the long-known concept of frequency shift keying (FSK) radar \cite{Barlow.1949} with microwave near-field imaging techniques, computational burden for the hand surface reconstruction is reduced by a factor of 1000, enhancing its suitability for real-time applications. In our previous work, we used only two closely neighbored frequencies to roughly localize the hand and estimate surface coordinates. This approach resulted in a high unambiguous range but low phase sensitivity regarding range measurements, impacting the accuracy of hand surface reconstruction. To enhance measurement accuracy and precision, while maintaining efficient measurement acquisition and reconstruction times, this work introduces a modified approach. The contributions of this paper are as follows:
\begin{figure}[t!]
	\centering
	\vspace{-2.5mm}
	\subfloat[]{\includegraphics[width=0.7357\columnwidth]{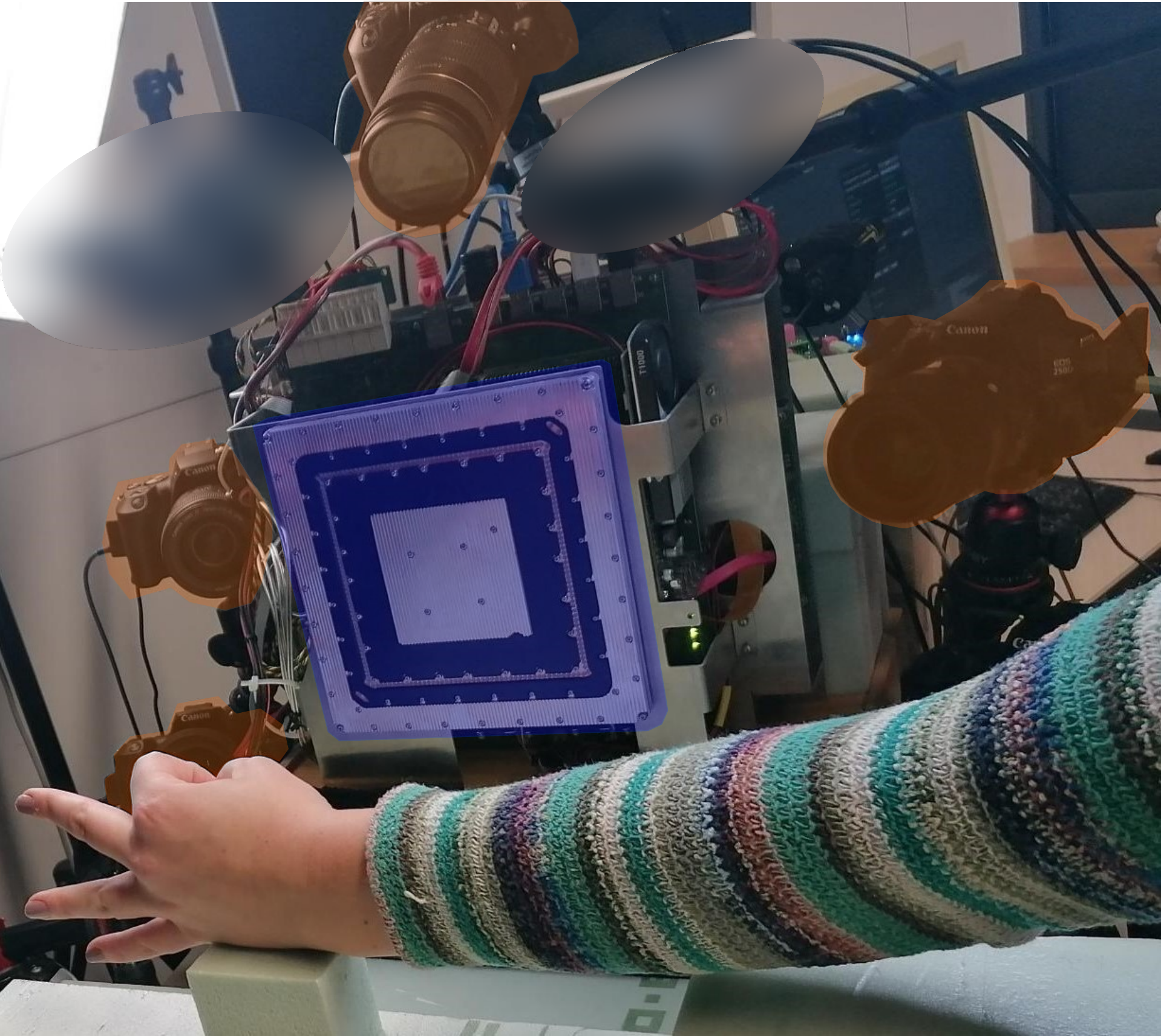}\label{fig:2a}}\hfill
	\subfloat[]{\includegraphics[width=0.1179\columnwidth]{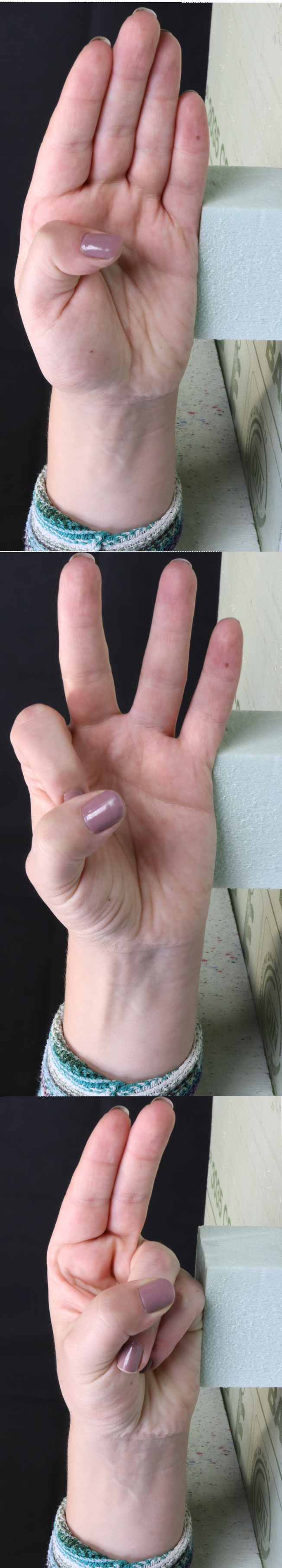}\label{fig:2b}}\hfill
	\subfloat[]{\includegraphics[width=0.1192\columnwidth]{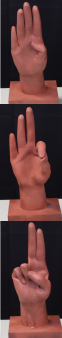}\label{fig_2c}}
	\caption{Measurement setup: (a) Imaging MIMO radar front-end (blue) comprising 94 Tx and 94 Rx antennas, photogrammetry setup consisting of multiple DSLR cameras (orange) and one exemplary hand pose placed on a styrodur element at a distance of approximately \SI{0.3}{\meter}, (b) hand poses from the American Sign Language alphabet utilized for accuracy evaluation, (c) 3D-printed hand poses utilized to evaluate precision.}
	\label{fig:2}
\end{figure}
\begin{itemize}[label=\textbullet, leftmargin=0.3cm]
\item We enhance the performance of the 3D hand imaging method from \cite{Braunig.2023} by transitioning from a two-tone (2-FSK) to a three-tone (3-FSK) radar technique.
\item We assess the accuracy of the 2-FSK, the 3-FSK, and a SOTA method utilizing the measurement setup seen in \autoref{fig:2}(a). The employed MIMO radar consists of 94 transmitting (Tx) and 94 receiving (Rx) antennas. The radar-based imaging results of the American Sign Language (ASL) alphabet hand poses depicted in \autoref{fig:2}(b) are compared to a ground truth derived from a spatially calibrated photogrammetry setup formed by multiple digital single-lens reflex cameras (DSLRs). 
\item The precision of both FSK methods is evaluated utilizing the 3D-printed hand poses depicted in \autoref{fig:2}(c).
\end{itemize}

\section{3-FSK Radar-based Hand Surface Imaging}\label{sec:3fsk}
\begin{figure}
	\centering
	\includegraphics[width=\columnwidth]{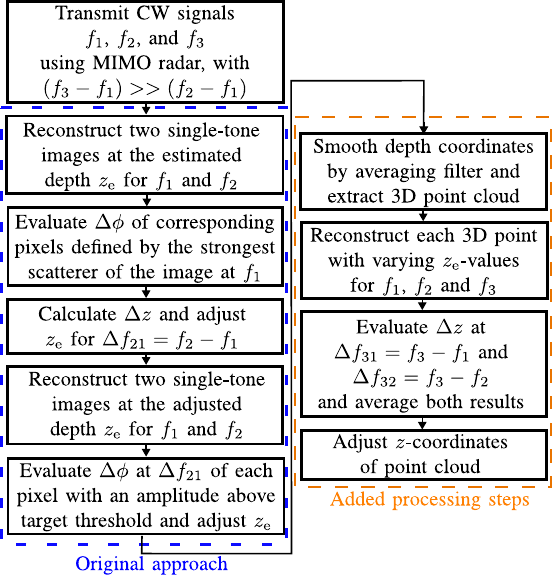} 
	\caption{Complete flowchart of the 3-FSK-based hand imaging concept.}
	\label{fig:flowchart}
\end{figure}
This work utilizes the method presented in \cite{Braunig.2023} to extract hand surface coordinates by evaluating the pixel-specific phase difference between two single-tone images $\hat{O}^{1/2}_{z_\mathrm{e}}$ measured by an imaging radar at two closely neighbored CW frequencies ($f_1$,$f_2$) and reconstructed at an estimated depth $z_\mathrm{e}$. The phase information $\Delta \phi = arg(\hat{O}_{z_\mathrm{e}}(x,y))$ of the complex difference pixel information, given by 
\begin{equation}
\hat{O}_{z_\mathrm{e}}(x,y) = \hat{O}_{z_\mathrm{e}}^{2}(x,y) \cdot \hat{O}_{z_\mathrm{e}}^{1}(x,y)^*
\end{equation} 
for the pixel at $(x,y,z_\mathrm{e})$, can be utilized to estimate the difference in depth $\Delta z = z - z_\mathrm{e}$ from the estimated point target position to the correct one at depth $z$.
The depth difference
\begin{equation}
	 \Delta z \approx \frac{c}{4\pi\Delta f}\Delta \phi
\end{equation}
is determined by the phase information $\Delta \phi$, where $c$ represents the speed of light, and $\Delta f = f_2 - f_1$ is the frequency difference between the CW signals. This is an iterative process. First, $\Delta z$ is evaluated for the pixel containing the highest amplitude. Subsequently, $z_\mathrm{e}$ is adjusted, two new single-tone images are reconstructed at the adjusted value, and only then $\Delta z$ is evaluated for every pixel above a predefined threshold, which yields an estimation for the $z$-coordinate of the hand surface within the respective pixels. As the maximum unambiguous deviation $\Delta z_\mathrm{max}$ is inversely proportional to $\Delta f$, a small value of $\Delta f = \SI{200}{\mega\hertz}$ is chosen to allow initial localization of the hand. 

The value of $\Delta f$ influences $\Delta z_\mathrm{max}$ but also the phase sensitivity regarding range $\sigma_{\Delta z}$ \cite{Zhang.2004}, which in this context impacts the smallest detectable value for $\Delta z$. Hence, the relationship
\begin{equation}
\sigma_{\Delta z} \sim \frac{360^{\circ}}{\Delta z_\mathrm{max}}
\end{equation}
is given. This means that for large values of $\Delta z_\mathrm{max}$ or small values for $\Delta f$, noise or other undesired influences, such as clutter and sidelobes, have a comparably high impact.

Therefore, we propose an adapted method that leverages the benefits of one small and one large frequency step. We employ a small $\Delta f$ to initially localize the hand and then roughly estimate surface coordinates for pixels above a certain threshold, likely containing the target. Afterwards, this data is smoothed by a 2D averaging filter, and a point cloud with varying values for $z_\mathrm{e}$ is extracted. Each 3D point within the point cloud is then reconstructed for all transmit frequencies $f_1$, $f_2$, and $f_3$, with $(f_3-f_1)>>(f_2~-~f_1)$. Subsequently, $\Delta z$ can be evaluated using the two large frequency differences $\Delta f_{31} = f_3-f_1$ and $\Delta f_{32}=f_3-f_2$. The respective values for $\Delta z$ are then averaged, yielding a correction for the estimated $z$-coordinates of the hand surface with higher accuracy and precision. Lastly, this 3D point cloud is smoothed by an averaging filter. A flowchart of the adapted concept can be seen in \autoref{fig:flowchart}. With only three CW signals being transmitted, frame rates can still be significantly increased compared to the SOTA. Further, reconstruction times are kept low. Compared to the original approach, the added number of calculation steps depends on the number of pixels containing a target $N_\mathrm{T}$. Hence, the additional computational effort consists of $3 \times N_\mathrm{T}$ calculation steps for the backprojection of the point targets and $2 \times N_\mathrm{T}$ steps for the evaluation of $\Delta \phi_{31}$ and $\Delta \phi_{32}$. Afterwards, the final $z$-values can be adjusted.
\begin{figure*}[t!]
	\centering
	\subfloat[]{\includegraphics[width=0.155\textwidth]{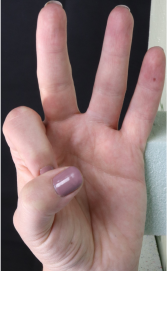}\label{figure:3foto}}\hfill
	\subfloat[]{\includegraphics[width=0.22\textwidth]{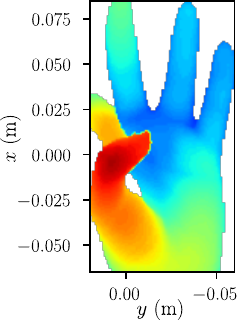}\label{figure:3a}}\hfill
	\subfloat[]{\includegraphics[width=0.1372\textwidth]{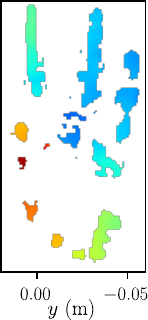}\label{figure:3b}}\hfill
	\subfloat[]{\includegraphics[width=0.1372\textwidth]{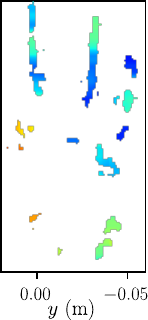}\label{figure:3c}}\hfill
	\subfloat[]{\includegraphics[width=0.22\textwidth]{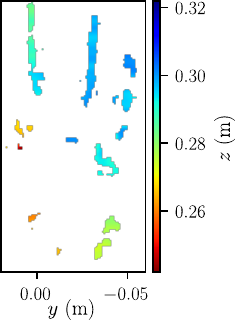}\label{figure:3d}}
	\caption{Comparison of the hand surface reconstructions for hand pose F. (a) Photo of the hand pose as well as (b) ground truth derived from photogrammetry in comparison to the respective reconstruction results using radar-based approaches: (c) broadband SOTA, (d) 2-FSK, and (e) 3-FSK.}
	\label{fig:3ges}
\end{figure*}
\section{Measurement Setup and Parameters}\label{sec:signal}

To evaluate the performance of the adapted approach, we performed multiple measurements using the setup visible in \autoref{fig:2}(a). The MIMO radar is a submodule of a commercially available automotive radome tester \cite{RohdeSchwarz}. It consists of 94 Tx and 94 Rx antennas and utilizes a stepped frequency continuous wave signal form. The length of the physical aperture is \SI{14}{\centi\meter}, resulting in a lateral resolution below \SI{5}{\milli\meter} at a \SI{30}{\centi\meter} distance \cite{Ahmed.2021b}. The distance between adjacent antennas is \SI{3}{\milli\meter}. The photogrammetry setup was spatially calibrated to the MIMO radar following the procedure from~\cite{Wirth.2024}. One of an author's hands was placed on a styrodur structure, around \SI{30}{\centi\meter} from the radar, and performed ASL alphabet hand poses (from \autoref{fig:2}(b)) to assess measurement accuracy. Additionally, we placed 3D-printed hand poses (from \autoref{fig:2}(c)) to establish a static object scene for precision evaluation.

The following signal and reconstruction parameters were chosen. For the 2-FSK-based approach, we chose $f_1=\SI{82}{\giga\hertz}$ and $f_2=\SI{81.8}{\giga\hertz}$, yielding $\Delta f_{21} = \SI{200}{\mega\hertz}$. In case of the 3-FSK-based approach the additional frequency was chosen at $f_3=\SI{79.5}{\giga\hertz}$, resulting in $\Delta f_{31} = \SI{2.5}{\giga\hertz}$ and $\Delta f_{32} = \SI{2.3}{\giga\hertz}$. Images were reconstructed for $x$- and $y$-values ranging from [\SI{-0.15}{\meter}, \SI{0.15}{\meter}] with a pixel dimension of $\SI{1}{\milli\meter} \times \SI{1}{\milli\meter}$. To define which pixel contains a target, we applied a threshold of $-\SI{10}{\decibel}$ below the maximum occurring amplitude. For the FSK-based approaches, we chose a first estimate for $z_\mathrm{e} = \SI{0.45}{\meter}$. When utilizing the broadband SOTA approach, $N_\mathrm{f} = 128$ frequency steps from \SI{72}{\giga\hertz} to \SI{82}{\giga\hertz} were used to reconstruct a 3D volume with the same sampling for $x$ and $y$ as the FSK-based approaches and a sampling in $z$ from \SI{0.245}{\meter} to \SI{0.325}{\meter} with a voxel dimension of $\SI{1}{\milli\meter} \times \SI{1}{\milli\meter} \times \SI{1}{\milli\meter}$. To extract the $z$-coordinate of the surface, a maximum intensity projection was performed, and a threshold of $\SI{-13}{\decibel}$ was applied. 

\section{Performance of Hand Surface Reconstruction}

To validate the proposed method, we evaluated its performance regarding the accuracy and precision of hand surface reconstruction.
\subsection{Accuracy Metrics}
\begin{figure}[t!]
	\centering
	\subfloat[]{\includegraphics[width =0.45\columnwidth]{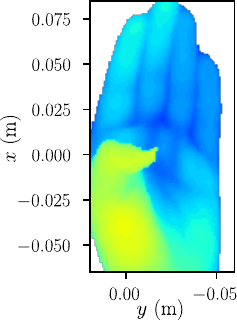}
		\label{figure:4a}}
	\subfloat[]{\includegraphics[width =0.45\columnwidth]{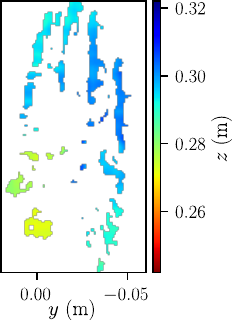}
		\label{figure:4b}}
	\vfill
	\subfloat[]{\includegraphics[width =0.45\columnwidth]{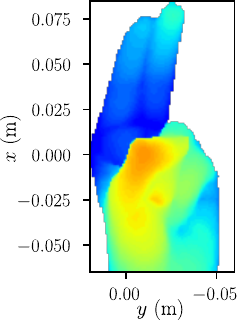}
		\label{figure:4c}}
	\subfloat[]{\includegraphics[width =0.45\columnwidth]{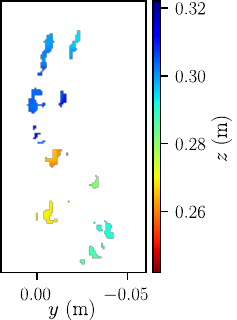}
		\label{figure:4d}}
	\caption{Hand surface reconstruction results using the 3-FSK-based approach (b) for hand pose B and (d) hand pose U in comparison to the respective ground truth derived from photogrammetry for hand pose (a) B and (c) U.}
	\label{fig:4ges2}
\end{figure}

As a measure to quantify accuracy, the absolute error, describing the difference of the estimated $z$-coordinate between the radar-based approaches and the outcome of the photogrammetry, was calculated for all ASL alphabet hand poses in \autoref{fig:2}(b). In \autoref{fig:3ges}, the ground truth hand surface generated from the photogrammetry as well as the radar-based reconstruction results for the adapted (3-FSK), the original (2-FSK), and the broadband SOTA method are displayed for hand pose F. Further, the reconstruction results for the adapted method in comparison to ground truth are shown in \autoref{fig:4ges2}. It should be noted that the absolute error can only be calculated for pixels that generate a signal amplitude above the respective threshold, which was \SI{-10}{\decibel} for the FSK-based approaches and \SI{-13}{\decibel} below the maximum amplitude for the broadband case. Since the hand surface predominantly produces specular reflections, some areas, in this case particularly the hand palm, exhibit a considerably low radar response owing to their unfavorable spatial orientation relative to the MIMO radar. \autoref{tab:accuracy} shows the mean absolute error, MAE, as well as the maximum absolute error, MAX of AE, averaged over all hand poses, when utilizing the 2-FSK, 3-FSK, and broadband SOTA approaches. The results show an improved value for both measures when comparing the original and the adapted approach. The MAE is reduced by approximately \SI{5}{\milli\meter}. With a value of \SI{1.7}{\milli\meter}, the 3-FSK approach achieves an accuracy that is comparable with the broadband SOTA approach that utilizes a bandwidth of \SI{10}{\giga\hertz}. This improvement is also visible when looking at \autoref{fig:3ges}. Despite fewer pixels containing an amplitude above the respective threshold in the FSK-based techniques compared to the broadband SOTA case (see \autoref{fig:3ges}(c)), the 3-FSK-based approach (see \autoref{fig:3ges}(e)) yields nearly identical results to the SOTA case and closely matches the respective ground truth (see \autoref{fig:3ges}(b)). However, when comparing the broadband SOTA method's outcome to the 2-FSK method (\autoref{fig:3ges}(d)), it inherits more deviations, clearly visible at the extended fingers. Furthermore, the 3-FSK-based reconstruction results for hand pose B and U visible in \autoref{fig:4ges2} also show great correspondence with the respective ground truth. By taking the mean of the absolute deviation in relation to the true $z$-value, the proposed approach yields an excellent accuracy of \SI{99.4}{\percent}.
For completeness, we also included accuracy metrics when utilizing only the 3-FSK frequencies and applying an SOTA backprojection with the same voxel size as in the broadband SOTA case, followed by a maximum intensity projection. The results are also depicted in \autoref{tab:accuracy} referenced as SOTA (narrowband). The 3-FSK method significantly outperforms the SOTA approach in accuracy, with an almost tenfold reduction in mean absolute error, under the assumption of equal bandwidth.

%
%
\begin{table}
	\caption{Performance metrics for reconstruction of the hand surface. The values are averaged for all three hand poses. MAE: mean absolute error; MAX of AE: maximum of absolute error.}
	\vspace{2pt}
	\centering
	\begin{tabular}{|p{2.2cm}|>{\raggedleft\arraybackslash}p{0.7cm}|>{\raggedleft\arraybackslash}p{0.7cm}|>{\raggedleft\arraybackslash}p{1.6cm}|>{\raggedleft\arraybackslash}p{1.6cm}|}
		\hline
		& \textbf{2-FSK} & \textbf{3-FSK} & \textbf{SOTA} (broadband) & \textbf{SOTA} (narrowband) \\
		\hline
		\textbf{MAE (mm)} & 6.6 & 1.7 & 1.0 & 13.4 \\
		\textbf{MAX of AE (mm)} & 39.2 & 28.8 & 32.0 & 64.9 \\
		\textbf{Precision (mm)} & 3.6 & 0.3 & - & -\\
		\hline
	\end{tabular}
	\label{tab:accuracy}
\end{table}
\subsection{Precision Metrics}
Furthermore, we assessed the precision of both FSK-based methods alongside their accuracy. To guarantee a completely static measurement setup, we positioned the three 3D-printed hand poses visible in \autoref{fig:2}(c), which were also used in \cite{Braunig.2023}, in front of the setup visible in \autoref{fig:2}(a). To calculate the precision, we evaluated the $z$-coordinate of the strongest scatterer for 10 measurements each and built the standard deviation of these values for each of the static hand poses. For the 2-FSK-based approach, this yields a precision of \SI{3.6}{\milli\meter}. With respect to the mean value, this gives a relative precision of \SI{98.3}{\percent}. Using the adapted approach, the precision can be enhanced to a value of \SI{0.3}{\milli\meter} or a relative value of \SI{99.99}{\percent}. These values were averaged over all printed hand poses.
	

\section{Conclusion}
In conclusion, this paper introduces a high-performance approach that extends the methodology outlined in \cite{Braunig.2023} for ultra-efficient 3D reconstruction of the hand surface. Transmitting three CW frequencies, the adapted concept exploits varied frequency differences for a high unambiguous range and high phase sensitivity. The method exhibits a remarkable reconstruction accuracy of \SI{99.4}{\percent}, as validated against a ground truth hand surface derived from a photogrammetry setup. It outperforms the 2-FSK-based method and even approaches the accuracy of a broadband SOTA method. Additionally, it achieves a precision of 0.3 millimeters (\SI{99.99}{\percent}), emphasizing its outstanding performance.
In summary, this method provides highly accurate and precise hand surface reconstruction crucial for hand shape and pose assessment, as well as tracking applications. Its reduced computational burden positions it favorably for real-time applications, such as radar-based gesture recognition. Moreover, transmitting only three CW frequencies allows for a substantial increase in frame rates compared to the broadband approach, thereby paving the way for simultaneous Doppler evaluations.

\section*{Acknowledgment}

The authors would like to thank the Rohde \& Schwarz GmbH \& Co. KG (Munich, Germany) for providing the radar imaging devices and technical support that made this work possible.
\par This work was funded by the Deutsche Forschungsgemeinschaft (DFG, German Research Foundation) -- SFB 1483 -- Project-ID 442419336, EmpkinS. 

\bibliographystyle{IEEEtran}
\bibliography{bibliography}

\begin{thebibliography}{1}
\providecommand{\url}[1]{#1}
\csname url@samestyle\endcsname
\providecommand{\newblock}{\relax}
\providecommand{\bibinfo}[2]{#2}
\providecommand{\BIBentrySTDinterwordspacing}{\spaceskip=0pt\relax}
\providecommand{\BIBentryALTinterwordstretchfactor}{4}
\providecommand{\BIBentryALTinterwordspacing}{\spaceskip=\fontdimen2\font plus
\BIBentryALTinterwordstretchfactor\fontdimen3\font minus
  \fontdimen4\font\relax}
\providecommand{\BIBforeignlanguage}[2]{{%
\expandafter\ifx\csname l@#1\endcsname\relax
\typeout{** WARNING: IEEEtran.bst: No hyphenation pattern has been}%
\typeout{** loaded for the language `#1'. Using the pattern for}%
\typeout{** the default language instead.}%
\else
\language=\csname l@#1\endcsname
\fi
#2}}
\providecommand{\BIBdecl}{\relax}
\BIBdecl

\bibitem{Huang.2021}
L.~Huang, B.~Zhang, Z.~Guo, Y.~Xiao, Z.~Cao, and J.~Yuan, ``Survey on depth and
  rgb image-based 3d hand shape and pose estimation,'' \emph{Virtual Reality \&
  Intell. Hardware}, vol.~3, no.~3, pp. 207--234, 2021.

\bibitem{Ahmed.2021}
S.~Ahmed, K.~D. Kallu, S.~Ahmed, and S.~H. Cho, ``Hand gestures recognition
  using radar sensors for human-computer-interaction: A review,'' \emph{Remote
  Sens.}, vol.~13, no.~3, p. 527, 2021.

\bibitem{Schuessler.2024}
C.~Schuessler, W.~Zhang, J.~Br{\"a}unig, M.~Hoffmann, M.~Stelzig, and
  M.~Vossiek, ``Radar-based recognition of static hand gestures in american
  sign language,'' 2024, arXiv:cs.CV/1107.1153.

\bibitem{Ahmed.2021b}
S.~S. Ahmed, ``Microwave imaging in security --- two decades of innovation,''
  \emph{IEEE J. Microw.}, vol.~1, no.~1, pp. 191--201, 2021.

\bibitem{Braunig.2023}
J.~Br{\"a}unig \emph{et~al.}, ``An ultra-efficient approach for high-resolution
  mimo radar imaging of human hand poses,'' \emph{IEEE Trans. Radar Syst.},
  vol.~1, pp. 468--480, 2023.

\bibitem{Barlow.1949}
E.~J. Barlow, ``Doppler radar,'' \emph{Proc. Inst. Radio Eng.}, vol.~37, no.~4,
  pp. 340--355, 1949.

\bibitem{Zhang.2004}
H.~Zhang and K.~Wu, ``Three-frequency principle for automotive radar system,''
  in \emph{Proc. IEEE Radio and Wirel. Conf.}, J.~S. Kennedy and G.~Heiter,
  Eds.\hskip 1em plus 0.5em minus 0.4em\relax Piscataway, NJ: {IEEE Operations
  Center}, 2004, pp. 315--318.

\bibitem{RohdeSchwarz}
\BIBentryALTinterwordspacing
``Rohde {\&} schwarz qar50 radome tester,'' 2022. [Online]. Available:
  \url{https://www.rohde-schwarz.com/qar50/}
\BIBentrySTDinterwordspacing

\bibitem{Wirth.2024}
V.~Wirth \emph{et~al.}, ``Automatic spatial calibration of near-field mimo
  radar with respect to optical sensors,'' 2024, arXiv:cs.RO/2403.10981.

\end{thebibliography}


\end{document}